\documentclass[journal]{IEEEtran}
\ifCLASSINFOpdf
\else
\fi
\usepackage{xcolor}
\usepackage{tabularx}
\usepackage{makecell}
\usepackage{graphicx}
\usepackage{subfig}
\usepackage{caption}
\usepackage[font=footnotesize]{caption}
\usepackage{amsmath}
\usepackage[switch,pagewise]{lineno} 
\usepackage{lipsum} 

\begin{document}
%
\title{Real-time Trust Prediction in Conditionally Automated Driving Using Physiological Measures}
%
%
%

\author{{Jackie Ayoub, Lilit Avetisian, X. Jessie Yang, Feng Zhou}
\thanks{J. Ayoub , L. Avetisian, and F. Zhou are with the Department of Industrial and Manufacturing, Systems Engineering, The University of Michigan, Dearborn, 4901 Evergreen Rd. Dearborn, MI 48128 USA (e-mail: \{jyayoub, lilita, fezhou\}@umich.edu).}
\thanks{X. J. Yang is with Industrial and Operations Engineering, University of Michigan, Ann Arbor, 500 S State St, Ann Arbor, MI 48109 USA (e-mail: xijyang@umich.edu).}
\thanks{Manuscript received Nov 17, 2022; revised xxx 26, 2022.}}

%
%

\markboth{IEEE Transactions on Intelligent Transportation Systems,~Vol.~xx, No.~xx, November~2022}%
{Ayoub \MakeLowercase{\textit{et al.}}: Predicting Takeover Time in Conditional Automated Driving}
%



\maketitle

\begin{abstract}
Trust calibration presents a main challenge during the interaction between drivers and automated vehicles (AVs). In order to calibrate trust, it is important to measure drivers' trust in real time. One possible method is through modeling its dynamic changes using machine learning models and physiological measures. In this paper, we proposed a technique based on machine learning models to predict drivers' dynamic trust in conditional AVs using physiological measurements in real time. We conducted the study in a driving simulator where participants were requested to take over control from automated driving in three conditions that included a control condition, a false alarm condition, a miss condition with eight takeover requests (TORs) in different scenarios. Drivers' physiological measures were recorded during the experiment, including galvanic skin response (GSR), heart rate (HR) indices, and eye-tracking metrics. Using five machine learning models, we found that  eXtreme Gradient Boosting (XGBoost) performed the best and was able to predict drivers' trust in real time with an f1-score of 89.1\%. Our findings provide good implications on how to design an in-vehicle trust monitoring system to calibrate drivers' trust to facilitate interaction between the driver and the AV in real time.
\end{abstract}

\begin{IEEEkeywords}
Trust prediction, physiological measures, real time,machine learning, automated vehicles.
\end{IEEEkeywords}

%
\IEEEpeerreviewmaketitle

\section{Introduction}
\IEEEPARstart{T}RUST is essential in shaping driver-AV (automated vehicle) interaction \cite{ayoub2021modeling, ayoub2021investigation}. 
On one hand, the public seems reluctant about using AVs due to a lack of trust \cite{menon2015consumer}. On the other hand, people have been spotted falling asleep behind the wheel of Tesla. To better manage driver-AV interaction, it is important for researchers to develop methods for real-time measures of a driver's trust in AVs.




A variety of techniques have been proposed to measure trust in real-time. The majority of previous studies have used self-reported assessment to measure trust (e.g., \cite{ayoub2021investigation, merritt2008not}).
Using such methods, participants are asked to rate their trust on a scale \cite{muir1987trust, chen2018planning}. These methods are simple, quick, and easy to implement with low costs \cite{mcdonald2008measuring}. However, implementing them in real world applications is challenging, as it is not practical to interrupt drivers to ask them to rate their trust repeatedly \cite{desai2013impact}. 

Behavioral measures of trust (e.g., reaction time and facial expressions) are an alternative approach to measuring trust in real time. They are helpful in assessing people's trust by observing and recording participants' behavioral processes \cite{korber2018introduction,miller2016behavioral}. For instance, Korber et al. \cite{korber2018introduction} used explicit (e.g., in the form of reliance) and implicit (e.g., in the form of eye tracking) behavioral measures of trust. Although eye tracking measures were useful in this task, reliance on the system was often specific to the task at hands and thus is often not well generalized to other situations. 

The third type of trust measures are physiological measures. They capture biological responses (e.g., hormonal levels, muscle movements, and neural activation) using electrocardiogram (ECG), electroencephalogram (EEG), eye gaze tracking, and skin conductance responses (SCRs) \cite{akash2018classification, de2018learning, lu2020modeling} during the human-machine interaction process. Although these techniques are correlated with human trust and do not interrupt drivers' tasks, there are generally no straightforward mapping relationships between these signals and trust dynamics in different contexts of human-machine interaction. 

In this paper, we investigated how system malfunctions affected participants' dynamic trust. In SAE Level 3 automated vehicles, the system was capable of identifying hazardous situations to trigger driver intervention. Two different malfunctions, i.e., misses and false alarms \cite{thropp2018calibrating}, were designed to elicit different levels of trust. Misses occurred when the system was not able to identify potential hazards. On the other hand, a false alarm occurred when the system identified non existing hazards. These different errors proved to have different impacts on trust in previous studies (e.g., \cite{azevedo2020real}).

In summary, the contributions of this study are as follows:
\begin{itemize}
  \item We used physiological measures to assess participants' trust in real-time in conditionally automated driving under two different system malfunctions (i.e., misses and false alarms).
  \item We built different machine learning models to predict trust dynamics in real time in conditionally automated driving and the XGBoost model had the best performance with an f1-score of 89.1\% . 
  \item The obtained results for the real-time prediction of trust provided important implications for designing trust-aware conditional automated vehicles. 

\end{itemize}

\section{Related Work}

\subsection{Factors influencing trust in automation and trust in AVs}

A wide range of factors have been identified as antecedents of people's trust in automation in general.  Schaefer et al. \cite{Schaefer:2016hp} categorized these factors into human-related, system-related, and environment-related factors. Bashir \cite{hoff2015} proposed another taxonomy and categorized these factors based on three types of trust on which a factor would have an effect, namely dispositional, learned, and situational trust \cite{hoff2015}.  Dispositional trust represents people's tendency to trust automation; learned trust is affected by people's past experience in using the automation system, and situational trust is dependent on the interaction between people and the automation in specific contexts. For a list of the factors influenceing trust in automation, please refere to \cite{Schaefer:2016hp, hoff2015}.

With the advances in the development of AVs, there is an increasing amount of research attention on trust in AVs \cite{ayoub2019}. For example, Ayoub et al. \cite{ayoub2021modeling} identified the factors (i.e., perceived benefits, perceived risks, excitement, knowledge, and the eagerness to adopt a technology) affecting dispositional and learned trust using machine learning models. Zhang et al. \cite{zhang2019} found that perceived risk was negatively correlated with initial learned trust while perceived usefulness was positively correlated with initial learned trust.
With respect to AVs, Zmud et al. \cite{zmud2016} reported that safety risks due to system failures dramatically influenced participants' trust in AVs, which further influenced their adoption of AVs. Dzindolet et al. \cite{dzindolet2003} showed that a transparent system that provided useful feedback and information about why a failure occured increased participants' trust in AVs. Luo et al. \cite{luo2020trust} showed that participants' trust decreased more with AVs' internal errors (e.g., sensor errors) compared to external errors (e.g., roadblocks). 
Okamura et al. \cite{okamura2020} calibrated participants' dynamic situational trust using calibration cues. For a list of factors influencing trust in AVs, please refer to \cite{ayoub2019}.

\subsection{Predicting trust in real-time}

To objectively measure trust in real-time, researchers argued that psychophysiological measurements, such as galvanic skin response (GSR), eye tracking, and heart rate are promising \cite{hergeth2016, walker2019, akash2018}. Eye tracking measures have been widely used as a real-time indicator of cognitive processes (i.e., situational awareness, decision making, and workload) \cite{vachon2014eye, zhou2021using}. They offer many advantages, including non-invasive, unobtrusive, and easy setup. Recently, researchers have shown that eye tracking data is promising to infer trust in real time \cite{hergeth2016keep, lu2019eye}. For instance, a lower monitoring frequency of the road was associated with higher trust in an AV \cite{hergeth2016}. 

GSR has been widely used in measuring anxiety, cognitive workload, and emotions \cite{hergeth2016, zhou2011affect,hu2016real}. GSR is a measure of the sweat-gland activity that captures arousal based on the skin conductivity. Researchers found that  GSR was correlated with trust. For instance, Khawaji et al. \cite{khawaji2015using} found that average GSR peaks and values were significantly affected by trust and cognitive workload. Walker et al. \cite{walker2019} showed that the higher the trust in the system, the more attention they paid to secondary tasks and the less they checked the road, and the lower the GSR, and combining GSR with gaze behavior led to a better prediction of trust. Akash et al. \cite{akash2018classification} used electroencephalography (EEG) and galvanic skin response (GSR) and a quadratic discriminant classifier to predict trust in AVs. The obtained prediction accuracy was 78.6\%.

Heart rate and heart rate variability are cardiovascular measures shown to be highly correlated with workload, risk, and drowsiness \cite{held2021heart, zhou2022predicting, zhou2020driver}. Heart rate variability indicates the degree to which the nervous system adapts to environmental changes through a cardiovascular regulation \cite{thayer2005psychosomatics}. During stressful situations, the sympathetic part of the autonomic nervous system is activated which results in a sudden increase of the heart activity. HR is defined as the number of times the heart beats per min and HRV is the time fluctuations between sequence of successive heart beats. To date, the relationships between HR, HRV and trust have not been investigated.

\section{Methodology}
\begin{figure*}[h!]
\centering
\includegraphics[width=.9\linewidth]{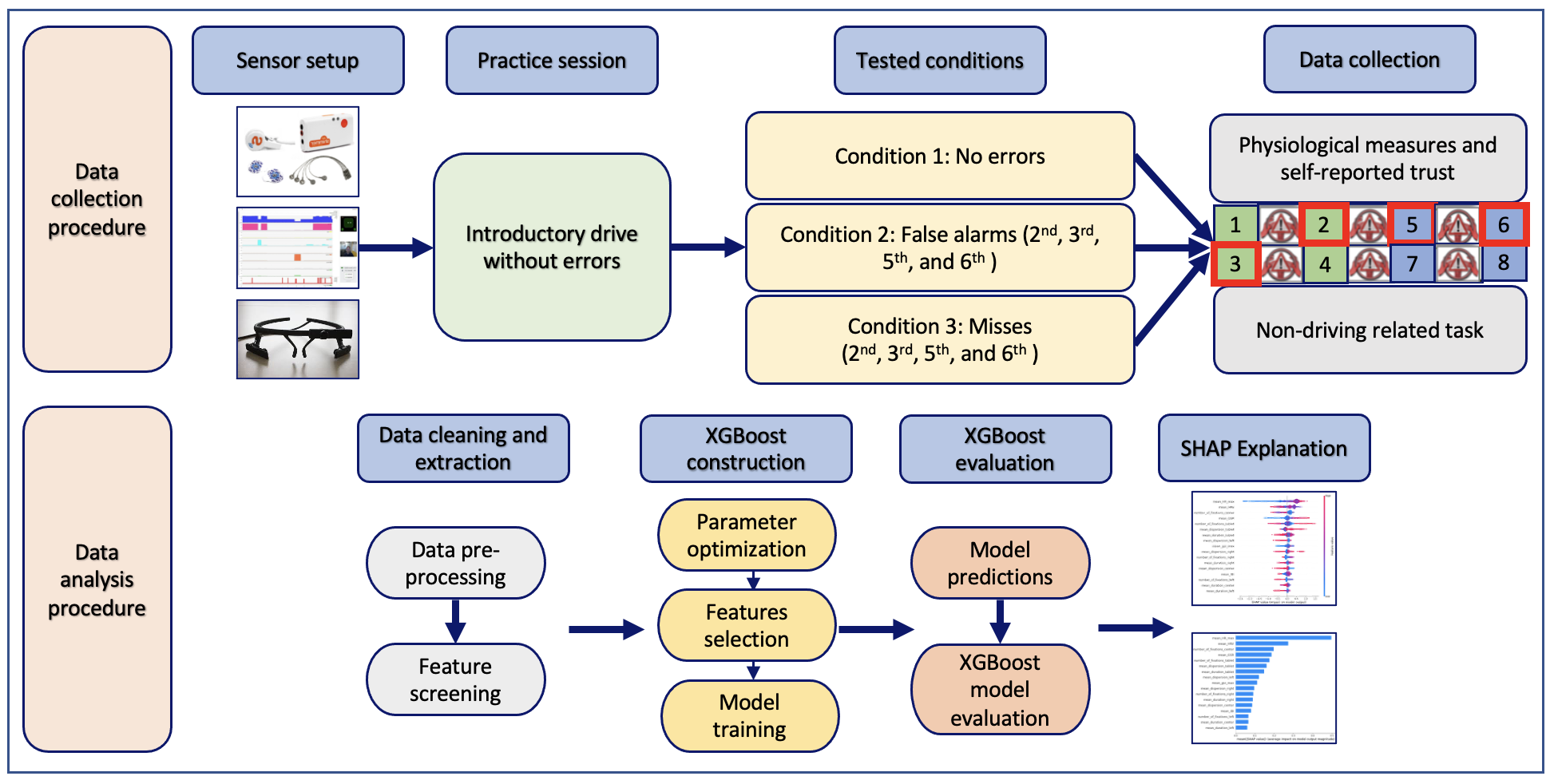}\hfill
\caption{Framework of the proposed study.}
\label{fig:expProc}
\end{figure*}

Fig. \ref{fig:expProc} provides an overview of the data collection and data analysis procedure.

\subsection{Participants} 
A total of 74 university students participated in this study. Due to malfunction of physiological sensors and the driving simulator, 15 participants were excluded and data from the remaining 59 participants (mean age = 21.3, SD = 2.9; ranging from 18 to 33; 26 females and 33 males) were used for further analysis (see Table \ref{tab:my-table1}). All the participants had a valid driver’s license and had normal or corrected-to-normal vision.  Participants received \$25 in compensation for about an hour of participation. The study was approved by the Institutional Review Board at the University of Michigan. 

\begin{figure}[tb!]
\centering
\includegraphics[width=.9\linewidth]{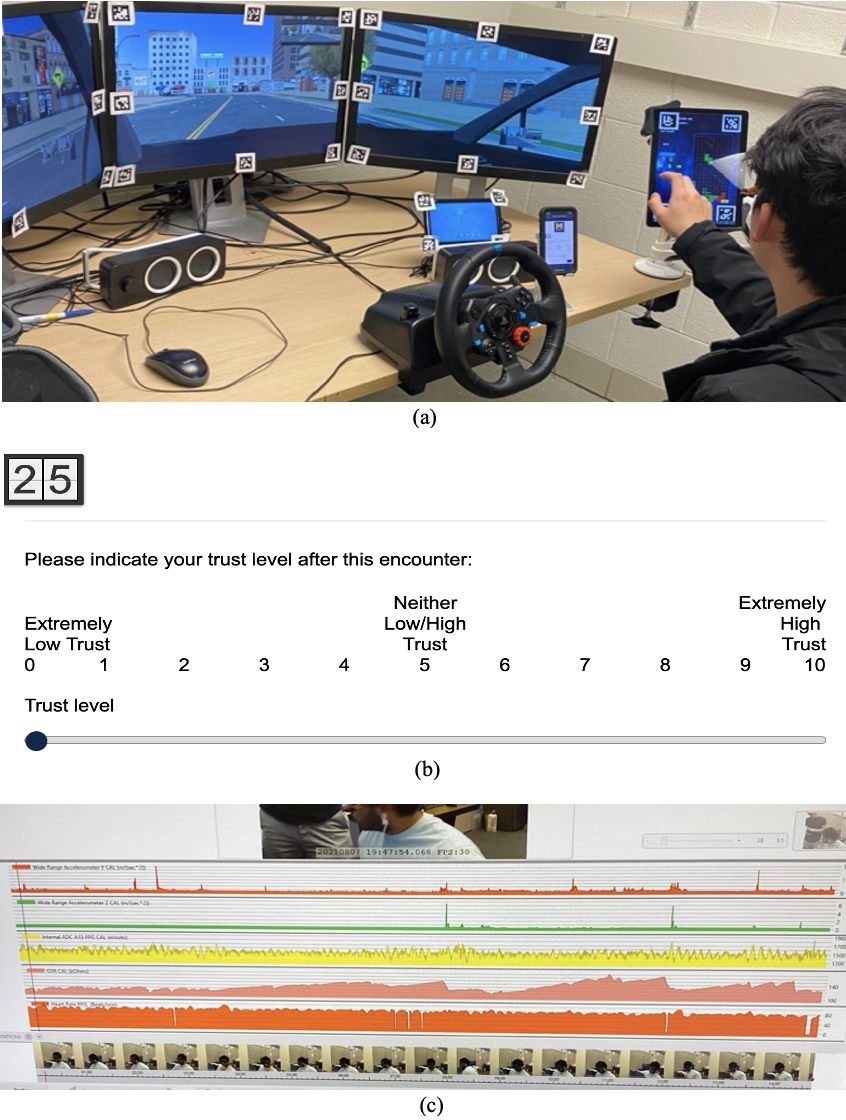}\hfill
\caption{(a) Experiment setup (b) Trust change self-report question (c) iMotion software.}
\label{fig:setup}
\end{figure}

\subsection{Apparatus and stimuli}
The study was conducted using a desktop-based driving simulator from Realtime Technologies Inc. (RTI, MI, USA). As shown in Fig. \ref{fig:setup}a, the simulator setup was composed of three LCD monitors integrated with a Logitech driving kit. Two other touchscreens (i.e., tablet and phone) were positioned to the right side of aparticipant for performing an non-driving related task (NDRT) and entering trust ratings. The NDRT was a Tetris game coded using PyGame library in Python. The game was designed in a way that the participants were required to drag the tiles otherwise they did not move. With this design, participants were able to pause the game to handle the TOR and get back to where they were before the TOR. A questionnaire designed in Qualtrics (Provo, UT, www.qualtrics.com) was used to evaluate participants’ trust. A single item question (``how much do you trust the AV?") was prompted on the screen every 25 seconds asking participants to rate their trust on a scale from 0 to 10 (see Fig. \ref{fig:setup}b). Prompting every 25 seconds was designed bench marking prior studies \cite{desai2013}.  
\begin{table}[]
\caption{Distribution of participants under the different conditions}
\label{tab:my-table1}
\begin{tabular}{cccccc}
\hline
\multicolumn{2}{c}{Control} & \multicolumn{2}{c}{FA} & \multicolumn{2}{c}{Misses} \\ \hline
Urban       & Suburban      & Urban    & Suburban    & Urban      & Suburban      \\
16          & 16            & 22       & 22          & 21         & 21            \\ \hline
\end{tabular}
\end{table}
The driving simulation was programmed to simulate SAE Level 3 automation and participants were able to control the vehicle laterally and longitudinally. To engage the automated mode, participants needed to press a red button on the steering wheel. Once the automated mode was engaged, participants would hear an auditory message, “Automated mode engaged”. Whenever a TOR was issued, participants would hear an auditory alert ``Takeover” and the automated mode would be automatically disengaged for the participant to take over control. If the participant did not take over control within the time limit, an auditory emergency stop warning (“Emergency Stop”) would be issued to avoid any crash. 

The Pupil Core (Pupil Lab, MA, USA) eye tracker headset was used to capture participants’ eye tracking data. The sampling rate of the eye-tracking system was 15 Hz. The Shimmer3 GSR+ Unit (Shimmer, MA, USA) was used to measure skin conductance and to photoplethysmogram (PPG) (used to calculate HR) data. Data from the Shimmer Unit was collected at a sampling rate of 128 Hz. The iMotions software (iMotions, MA, USA) was used for physiological data synchronization in real-time (see Fig. \ref{fig:setup}c).

\subsection{Experimental design}
In this study, participants were required to take over from automated driving eight times. The experiment was a one-way between-subjects design, in which the participants were randomly assigned to three conditions: 1) the control condition, i.e., all the eight TORs were true alarms, 2) false alarms, i.e., of the eight TORs, the 2nd, 3rd, 5th, and 6th were false alarms, and 3) misses, i.e., of the eight TORs, the 2nd, 3rd, 5th, and 6th were misses. The purpose of this design was to elicit different levels of trust since it was shown that both misses and false alarms degraded operator trust in automation \cite{pop2015individual}.

\begin{figure}[tb!]
\centering
\includegraphics[width=.9\linewidth]{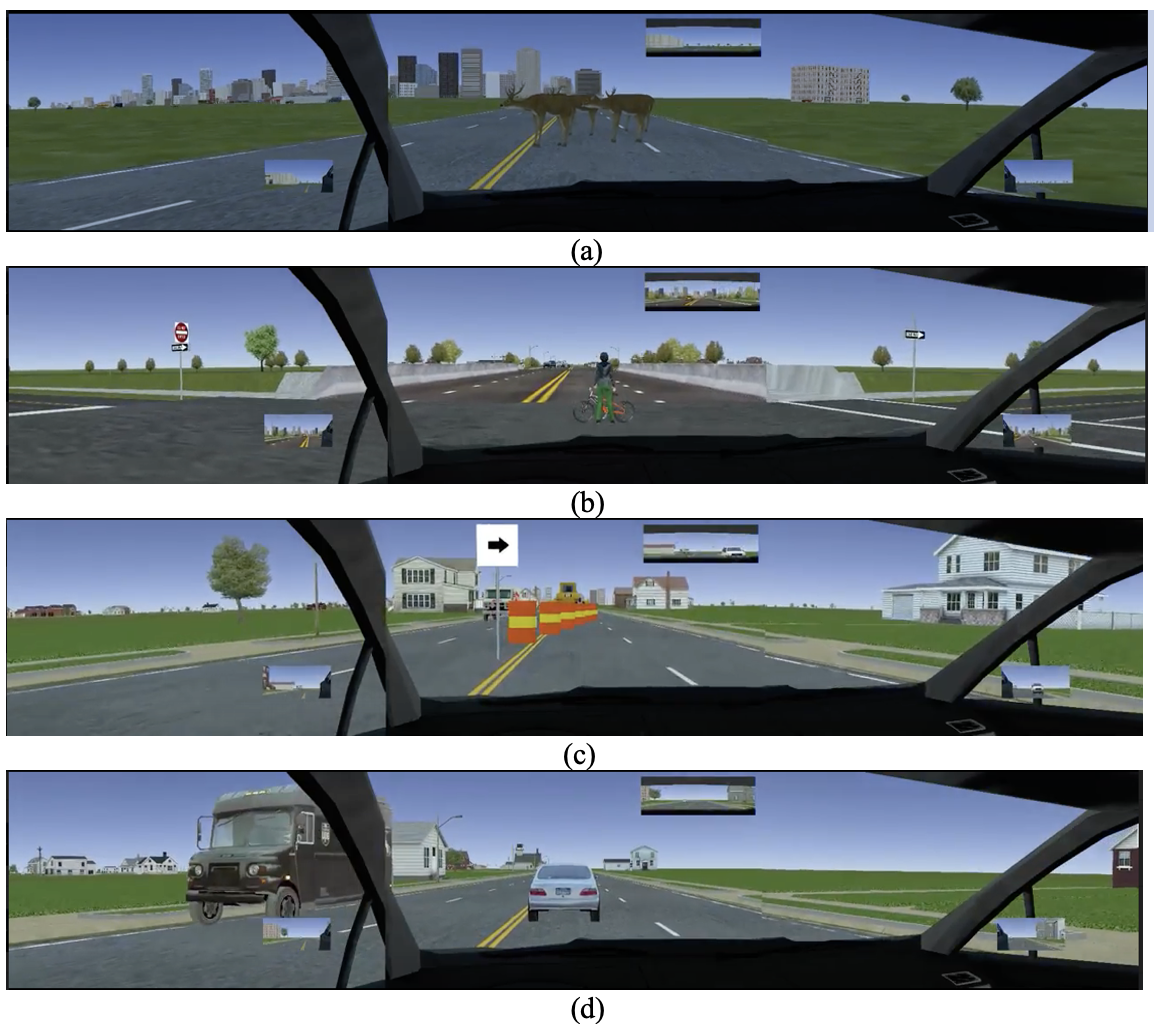}\hfill
\caption{Takeover events in suburban areas (a) dears ahead (b) bicyclist crossing ahead (c) construction zone ahead (d) vehicle sudden stop ahead.}
\label{fig:events1}
\end{figure}
\begin{figure}[tb!]
\centering
\includegraphics[width=.9\linewidth]{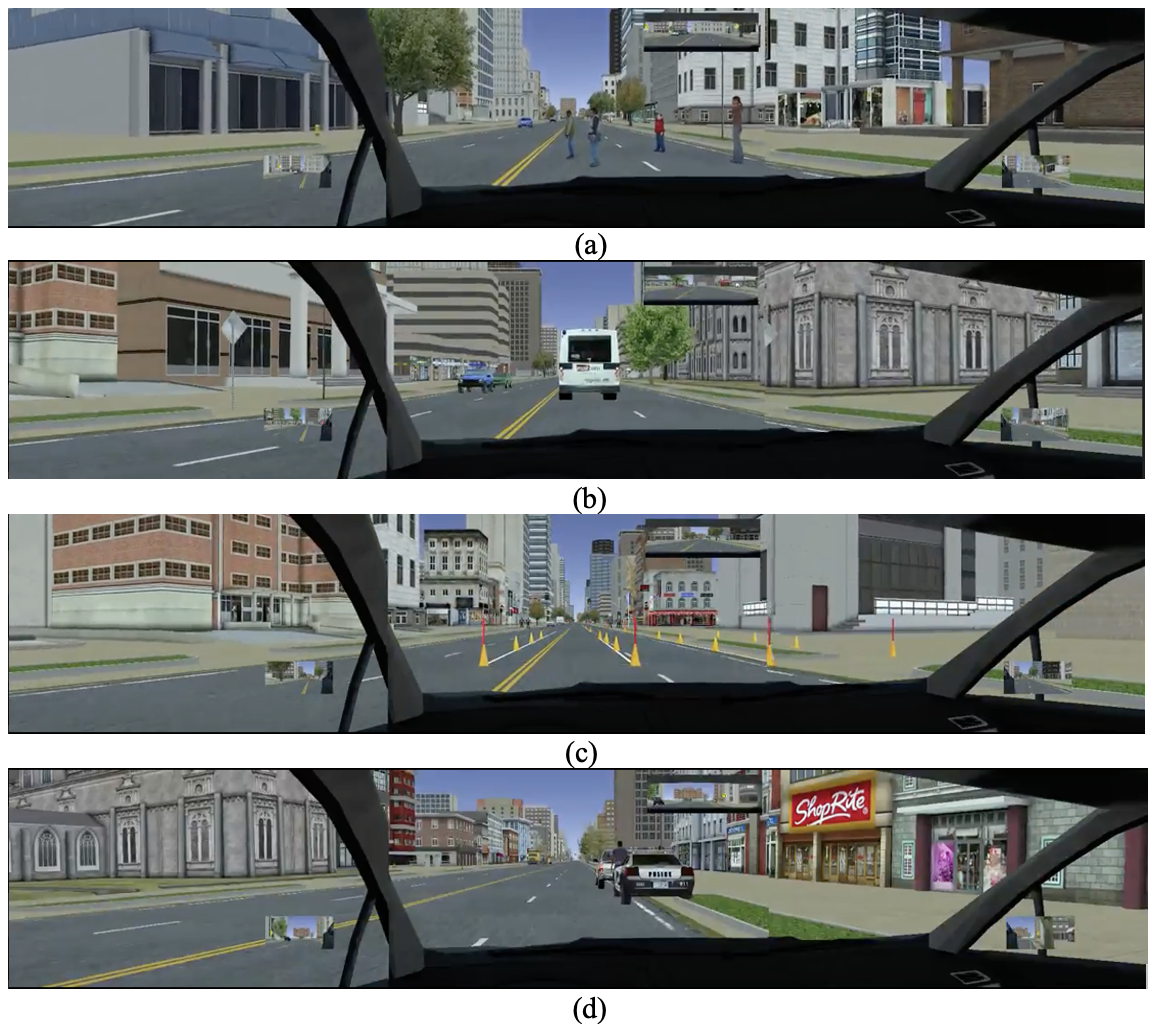}\hfill
\caption{Takeover events in urban areas (a) pedestrians crossing ahead (b) bus sudden stop ahead (c) construction zone ahead (d) police vehicle on shoulder.}
\label{fig:events2}
\end{figure}

Typical roadway features were used when a TOR was issued (i.e., (1) deer ahead, (2) bicyclist crossing ahead, (3) construction zone ahead, (4) vehicle sudden stop ahead, (5) pedestrians crossing ahead, (6) bus sudden stop ahead, (7) construction zone ahead, and (8) police vehicle on shoulder) (See Fig. \ref{fig:events1} and Fig. \ref{fig:events2}). In addition, four takeover scenarios happened in rural areas and the other four in urban areas. The order of urban and rural takeover scenarios was counterbalanced. 


\subsection{Experimental procedure}
Upon arrival, participants were asked to complete a consent form and an online demographic survey. After the survey, participants received an introduction and watcheda short video about the tasks they needed to do. Participants then completed a training session to familiarize themselves with the driving simulator and the experiment flow. Participants were informed that the vehicle would be able to perform the driving task when the automated mode was engaged, but they had to stay alert and be ready to take over whenever requested. 
We further explained that the AV could fail to detect obstacles for the participants to go through the miss condition and that the AV could give false alarms of TORs for those to go through the FA condition. 
Next, we calibrated the eye-tracker device by asking the participant to look at ten targets on the front screens. 
After that, we attached the GSR electrodes to the left foot of the participants and the PPG probe to their left ear lobe. Each drive (i.e., urban or suburban) took around 15 minutes and the whole experiment lasted around 75 minutes. In order to create the ground truth of trust prediction model, the participants were asked to respond to a single-item trust prompt on a scale from 0 to 10, “Please indicate your trust level after this encounter” \cite{hergeth2016}. Following Desai et al. \cite{desai2013}, participants were prompted for this trust measure every 25 seconds to ensure that they were not overwhelmed.

\subsection{Comparison between the three tested conditions}
A one-way analysis of variance (ANOVA) showed a significant difference in trust among the three conditions (i.e., control, misses, and FA), $F(2,47) = 22.323, p < 0.001$. A Tukey post-hoc test revealed that trust was significantly higher in the control ($M = 7.967$) and FA ($M = 7.699$) conditions compared to the misses condition ($ M = 5.597$) ($p < 0.001$). There was no significant difference between the FA and the control condition. Since the FA condition did not reduce participants’ trust, we did not include it in the trust prediction model in order to better balance two levels of trust in the dataset for machine learning. 

\section{Trust prediction model development}
Physiological (i.e., galvanic skin response, heart rate, and eye-tracking metrics) and self-reported data were collected for the trust model development. To train the trust prediction model, 17 features were extracted from the data (see Table 2). A five-fold cross-validation was used to optimize the f1-score of the prediction using a randomized search for hyperparameters.

\subsection{Data pre-processing}
The GSR is composed of phasic (i.e., fast variation of skin conductance) and tonic (slow variation of skin conductance) phases. In this study, the phasic component was used since it captures the GSR changes in seconds. Therefore, we used a continuous decomposition technique (i.e., Ledalab in MATLAB) \cite{benedek2010continuous}. The iMotion software was used to extract the heart rate related measures from the inter-beat interval. For eye-tracking data, the Pupil Player software was used for exporting the data collected in Pupil Core for further analysis.

\subsection{Model features}
As shown in the previous sections, our data was collected using different sensors (i.e., Shimmer and eye-tracking) and systems (i.e., Qualtrics). Therefore, we synchronized the time between GSR, HR, eye-tracking, and continuous trust data using the timestamps associated with them. After synchronization, we used a sliding window of 25 seconds before the self-reported trust level to extract a series of GSR, HR, and eye-tracking values within that window. Therefore, each self-reported rating was associated with a series of GSR, HR, and eye-tracking values at that corresponding timestamp. The extracted 17 features are listed in Table \ref{tab:my-table2} based on our previous studies \cite{du2020psychophysiological,du2020predicting}.


\begin{table*}
  \centering
  \small
  \caption{Description of the generated features}
\begin{tabularx}{\textwidth}{ll}

\hline
Physiological Data & Model Features \\ 
\hline
Heart rate & Heart rate max, heart rate variability, inter-beat interval\\

Fixation &  Number of fixations on the center, left, right, and NDRT screens\\

Duration &  Duration of fixations on the center, left, right, and NDRT screens\\

Dispersion &  Distance between all gaze locations during a fixation on the center, left, right, and NDRT screens\\

Galvanic skin response &  Mean and max of galvanic skin response in phasic phase\\

\hline
\end{tabularx}
\label{tab:my-table2}
\end{table*}

\subsection{Model development}
The trust prediction model was trained with an XGBoost model \cite{chen2016xgboost} for the following reasons. First, XGBoost is a decision tree model that combines multiple trees where each decision tree learns from the previous one to build a robust model. Second, XGBoost has the power to perform parallel processing and the advantage to improve the learning process without overfitting. In addition, XGBoost works with missing data without affecting the model performance significantly. Third, the XGBoost model can provide feature importance and help in interpreting the model predictions with the usage of SHapley Additive exPlanations (SHAP) explainer \cite{ayoub2022predicting}. The XGBoost model performance was also compared with other algorithms (e.g., logistic regression (LR), decision tree (DT), Naïve Bayes (NB), and K-nearest neighbors (KNN)). The response variable was defined as a binary one (i.e., trust = 1 (trust value $>=$ 5, sample size = 1745) and distrust = 0 (trust value $<$ 5), sample size = 484). The objective function used was binary: logistic. The performance metrics used to evaluate the XGBoost model were accuracy, f1-score, precision, recall, and ROC\_AUC. SHAP explainer was used to explain the predictions made by the XGBoost model. Specifically, SHAP explainer helped in ranking the features based on their importance.

\section{Results}

\subsection{XGBoost performance}
The XGBoost model performance is shown in Table \ref{tab:my-table3} using 10-fold cross validation. Also, 10-fold cross validation was used to compare the XGBoost performance with other algorithms (e.g., LR, DT, NB, and KNN). As shown in Table \ref{tab:my-table3}, XGBoost performed the best almost across all metrics (i.e., accuracy, f1-score) among the list of machine learning models. 
Since the dataset was imbalanced, we varied the sample size of the trust data as shown in Table \ref{tab:my-table4}. In the first trial, we randomly selected the same number of sample size for the trust and distrust data. In the second trial, we doubled the sample size of trust data randomly. And in the third trial, we tripled the sample size of trust data randomly. A 10-fold cross validation process was used to obtain the results shown in Table \ref{tab:my-table4}. The XGBoost model had better performance with the increase in the sample size of trust data. 

\begin{table}[]
\caption{Performance measure comparison between different models.}
\label{tab:my-table3}
\resizebox{\columnwidth}{!}{%
\begin{tabular}{lllll}
\hline
\textbf{Models}  & \textbf{Accuracy} & \textbf{Precision} & \textbf{Recall} & \textbf{f1-score} \\ \hline
Logistic Regression & 79.6\% & 79.9\% & 98.9\% & 88.3\% \\
Decision Tree       & 75.7\% & 82.8\% & 87.1\% & 84.9\% \\
Naïve Bayes         & 75.1\% & 80.5\% & 89.9\% & 84.9\% \\
KNN                 & 75.6\% & 84.3\% & 84.6\% & 84.5\% \\
\textbf{XGBoost} & \textbf{81.6\%}    & 83.4\%              & 95.5\%           & \textbf{89.1\%}    \\ \hline
\end{tabular}%
}
\end{table}

\begin{table}[]
\caption{Summary of XGBoost classifier performance}
\label{tab:my-table4}
\resizebox{\columnwidth}{!}{%
\begin{tabular}{lllll}
\hline
\textbf{Sample size} & \textbf{Accuracy} & \textbf{Precision} & \textbf{Recall} & \textbf{f1-score} \\ \hline
Trust (484), Distrust (484)  & 74.8\% & 72.5\% & 79.9\% & 75.9\% \\
Trust (968), Distrust (484)  & 78.5\% & 79.3\% & 91.8\% & 85.0\% \\
Trust (1452), Distrust (484) & 79.8\% & 81.2\% & 94.9\% & 87.6\% \\
Trust (1745), Distrust (484) & 81.6\% & 83.4\% & 95.5\% & 89.1\% \\ \hline
\end{tabular}%
}
\end{table}

\begin{figure*}[ht!]
\centering
\includegraphics[width=.6\linewidth]{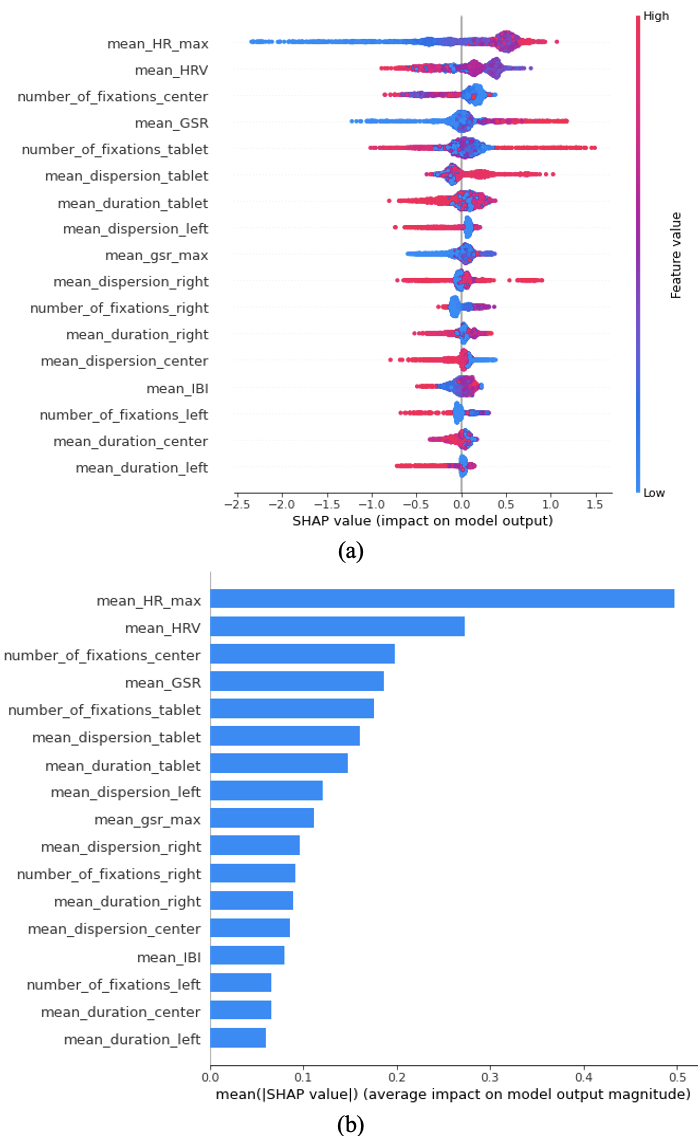}\hfill
\caption{(a) SHAP summary plot (b) SHAP feature importance plot.}
\label{fig:SHAP}
\end{figure*}

\subsection{Feature importance}
To understand the importance of each feature in predicting trust, SHAP feature importance plot was examined. Following our previous study \cite{ayoub2022cause, ayoub2022predicting}, we used feature selection to identify the best model performance by adding one feature at a time following the importance ranking of the variables identified in Fig. \ref{fig:SHAP}. We continued this process until the f1-score stopped improving. F1-score was used as our main performance measure since it is a harmonic mean of precision and recall and better than accuracy when the data is unbalanced for classification analysis. Finally, we found that XGBoost performed the best when a combination of seven features (i.e., 1) mean\_HR\_max, 2) mean\_HRV, 3) number\_of\_fixations\_center, 4) mean\_GSR, 5) number\_of\_fixations\_tablet, 6) mean\_dispersion\_tablet, and 7) mean\_duration\_tablet) was used. The SHAP summary plot shown in Fig. \ref{fig:SHAP}a has four characteristics, including 1) the density that represents the distribution of the features in the data, 2) the color that shows the range of a particular feature from high (red) to low (blue), 3) the horizontal variation that shows the large or small effect of the feature on the prediction, and 4) the vertical ranking that represents the importance of the feature. Take the feature “number\_of\_fixations\_center” as an example, a high number of fixations at the center screen led to a decrease in trust whereas a low number of fixations at the center screen led to an increase of trust.

\section{Discussion}

The purposes of this paper were: 1) to assess participants' trust in real-time in conditionally automated driving under two different system malfunctions (i.e., misses and false alarms) using physiological measures, 2) to build different machine learning models to predict trust dynamics in real time in conditionally automated driving, and 3) to provide implications for the design of trust-aware conditional automated vehicles. The results of the study are organized around two main findings. First, we found that XGBoost performed the best and was able to predict drivers' trust in real time with an f1-score of 89.1\%. Second, we identified the most important physiological measures for real-time prediction of trust. 


\subsection{Effects of features on trust prediction model}
In this study, we proposed a technique to use a combination of physiological measures, including heart rate activity, galvanic skin responses, and eye-tracking metrics to predict drivers’ dynamic trust. Previous studies have used physiological data to measure trust. For instance, Hergeth et al. \cite{hergeth2016} found a negative correlation between monitoring frequency and drivers’ self-reported trust measures. For the feature, “number\_of\_fixations\_center” in Fig. \ref{fig:SHAP}a, a high number of fixations (feature values in red) on the center screen decreased participants’ trust (SHAP value decreased) whereas a low number of fixations (feature values in blue) at the center screen increased their trust (SHAP value increased). 
As for the feature “number\_of\_fixation\_tablet”, its effect on trust was not clear as shown in Fig. \ref{fig:SHAP}a. Further investigation is needed to examine this.

For the heart rate related features, a high “mean\_HR\_max” increased participants’ trust while a low “mean\_HR\_max” decreased their trust. 
Researchers showed that heart rate increases were associated with stimulus ignorance and environmental rejection while heart rate decreases were associated with intake and enhancement of environmental stimuli \cite{du2020psychophysiological,libby1973pupillary}. It seemed reasonable that increased heart rate was associated with a higher level of trust that the participants could ignore the information during the drive and focus more on NDRTs and vice versa. This measure played an important role as it was the most important factor as shown in Fig. \ref{fig:SHAP}.
As for the “mean\_HRV” feature, a high “mean\_HRV” was associated with a lower level of trust while a low “mean\_HRV” was associated with a higher level of trust. A higher heart rate variability was reported to be associated with a lower mental workload \cite{du2020psychophysiological, zhou2022predicting}. This might be used to explain that those in the miss condition possibly reduced their mental workload as fewer TORs were issued. However, once they realized there were misses, they were more alert to be cognitively and behaviorally prepared, which indicated that a higher HRV was related to potentially better cognitive and behavioral scores \cite{liu2022heart} in a low trust condition. However, more studies are needed to further validate such explanations.

With respect to the “mean\_GSR” feature, a high “mean\_GSR” was associated with a high level of trust while a low “mean\_GSR” was associated with low trust. Usually a high GSR is associated with a high level of arousal \cite{zhou2011affect}, which might be contributed by the better gaming experience of Tetris of those in the control condition (a high level of trust) than those in the miss condition (a low level of trust). In a previous study, Khawaji et al. \cite{akash2018} showed that GSR values were significantly lower when both trust was high and cognitive load was low in a text-chat environment. Our finding was inconsistent with theirs, possibly because trust was coupled with cognitive load in their experiment. Thus, more studies are needed to resolve such inconsistencies. 

\subsection{Implications}

The research findings of this study have important implications for the design of automated driving systems. Trust is dynamically modeled and predicted using physiological measures. Such a representation of trust permits continuous capture of trust over time in automated driving. 
It thus provides important implications for the design of a trust calibration system. The system would help in calibrating the driver's trust by tracking the moments when the driver overtrust or undertrust the AV in real-time using physiological data and machine learning models. Calibration of trust to an appropriate level is considered a design goal to improve the safety and maximize the benefits of AVs while avoiding misuse/disuse of AVs \cite{ayoub2021investigation}. Thus, such a calibration system might help in improving people's acceptance and adoption of AV in the near future.

Moreover, the findings of this study would help advance our understanding of trust as a determining factor to optimize the interaction between the driver and the AV system. As shown in the result section, our proposed framework has good performance in trust predictions in real-time. However, better performance is needed to be applicable in reality.  Our proposed model used data of a group of users and in reality we can use this model as the base model and personalize it with individual drivers to further improve its performance.  In this sense, our research paves the way for trust prediction in real time using more complex models (such as deep learning models) with physiological data.

\subsection{Limitations}
Our study has limitations that are left for future investigations. First, the study was conducted in a low-fidelity experimental setup with a desktop driving simulator. Second, a sample size of 59 participants was considered small with respect to the three tested conditions (i.e., control, misses, and FA). In addition, the participants were mainly university students which led to a homogeneous sample in regards to age, education, driving experience, and knowledge about AVs. In the future, similar studies should be conducted in a higher fidelity setup and with a larger and diverse sample size. 
Third, trust was mainly evaluated in takeover scenarios in conditionl AVs using drivers' dynamic trust in control, misses, and FA conditions. In addition, the study covered a small number of possible takeover issues. Thus, future studies should explore trust in other scenarios, such as continuous driving performances in object detection, different types of driving styles, and recommendations on route selections, and so on.  The investigations that overcome such limitations mentioned above might offer a better understanding of the relationships between physiological measures and trust in automated driving.

\section{Conclusions}

In this study, we predicted drivers’ trust in real-time using physiological data using machine learning models in conditionally automated driving. Compared to previous studies, our proposed technique was shown to be effective in estimating drivers' trust. The XGBoost model had an accuracy of 81.6\% and an f1-score of 89.1\%, which outperformed other machine learning models. In addition, we identified the most important physiological measures for real-time prediction of trust, including 1) mean\_HR\_max, 2) mean\_HRV, 3) number\_of\_fixations\_center, 4) mean\_GSR, 5) number\_of\_fixations\_tablet, 6) mean\_dispersion\_tablet, and 7) mean\_duration\_tablet. Such a technique has important implications in the future to guide the design of an in-vehicle trust calibration system to improve people’s acceptance and trust in AVs. 

\section{Acknowledgement}
This research was supported by National Science Foundation.

\bibliography{main}

\begin{thebibliography}{10}
\providecommand{\url}[1]{#1}
\csname url@samestyle\endcsname
\providecommand{\newblock}{\relax}
\providecommand{\bibinfo}[2]{#2}
\providecommand{\BIBentrySTDinterwordspacing}{\spaceskip=0pt\relax}
\providecommand{\BIBentryALTinterwordstretchfactor}{4}
\providecommand{\BIBentryALTinterwordspacing}{\spaceskip=\fontdimen2\font plus
\BIBentryALTinterwordstretchfactor\fontdimen3\font minus
  \fontdimen4\font\relax}
\providecommand{\BIBforeignlanguage}[2]{{%
\expandafter\ifx\csname l@#1\endcsname\relax
\typeout{** WARNING: IEEEtran.bst: No hyphenation pattern has been}%
\typeout{** loaded for the language `#1'. Using the pattern for}%
\typeout{** the default language instead.}%
\else
\language=\csname l@#1\endcsname
\fi
#2}}
\providecommand{\BIBdecl}{\relax}
\BIBdecl

\bibitem{ayoub2021modeling}
J.~Ayoub, X.~J. Yang, and F.~Zhou, ``Modeling dispositional and initial learned
  trust in automated vehicles with predictability and explainability,''
  \emph{Transportation Research Part F: Traffic Psychology and Behaviour},
  vol.~77, pp. 102--116, 2021.

\bibitem{ayoub2021investigation}
J.~Ayoub, L.~Avetisyan, M.~Makki, and F.~Zhou, ``An investigation of drivers’
  dynamic situational trust in conditionally automated driving,'' \emph{IEEE
  Transactions on Human-Machine Systems}, vol.~52, no.~3, pp. 501--511, 2021.

\bibitem{menon2015consumer}
N.~Menon, \emph{Consumer perception and anticipated adoption of autonomous
  vehicle technology: Results from multi-population surveys}.\hskip 1em plus
  0.5em minus 0.4em\relax University of South Florida, 2015.

\bibitem{merritt2008not}
S.~M. Merritt and D.~R. Ilgen, ``Not all trust is created equal: Dispositional
  and history-based trust in human-automation interactions,'' \emph{Human
  factors}, vol.~50, no.~2, pp. 194--210, 2008.

\bibitem{muir1987trust}
B.~M. Muir, ``Trust between humans and machines, and the design of decision
  aids,'' \emph{International journal of man-machine studies}, vol.~27, no.
  5-6, pp. 527--539, 1987.

\bibitem{chen2018planning}
M.~Chen, S.~Nikolaidis, H.~Soh, D.~Hsu, and S.~Srinivasa, ``Planning with trust
  for human-robot collaboration,'' in \emph{Proceedings of the 2018 ACM/IEEE
  international conference on human-robot interaction}, 2018, pp. 307--315.

\bibitem{mcdonald2008measuring}
J.~D. McDonald, ``Measuring personality constructs: The advantages and
  disadvantages of self-reports, informant reports and behavioural
  assessments,'' \emph{Enquire}, vol.~1, no.~1, pp. 1--19, 2008.

\bibitem{desai2013impact}
M.~Desai, P.~Kaniarasu, M.~Medvedev, A.~Steinfeld, and H.~Yanco, ``Impact of
  robot failures and feedback on real-time trust,'' in \emph{2013 8th ACM/IEEE
  International Conference on Human-Robot Interaction (HRI)}.\hskip 1em plus
  0.5em minus 0.4em\relax IEEE, 2013, pp. 251--258.

\bibitem{korber2018introduction}
M.~K{\"o}rber, E.~Baseler, and K.~Bengler, ``Introduction matters: Manipulating
  trust in automation and reliance in automated driving,'' \emph{Applied
  ergonomics}, vol.~66, pp. 18--31, 2018.

\bibitem{miller2016behavioral}
D.~Miller, M.~Johns, B.~Mok, N.~Gowda, D.~Sirkin, K.~Lee, and W.~Ju,
  ``Behavioral measurement of trust in automation: the trust fall,'' in
  \emph{Proceedings of the Human Factors and Ergonomics Society Annual
  Meeting}, vol.~60, no.~1.\hskip 1em plus 0.5em minus 0.4em\relax SAGE
  Publications Sage CA: Los Angeles, CA, 2016, pp. 1849--1853.

\bibitem{akash2018classification}
K.~Akash, W.-L. Hu, N.~Jain, and T.~Reid, ``A classification model for sensing
  human trust in machines using eeg and gsr,'' \emph{ACM Transactions on
  Interactive Intelligent Systems (TiiS)}, vol.~8, no.~4, pp. 1--20, 2018.

\bibitem{de2018learning}
E.~J. De~Visser, P.~J. Beatty, J.~R. Estepp, S.~Kohn, A.~Abubshait, J.~R.
  Fedota, and C.~G. McDonald, ``Learning from the slips of others: Neural
  correlates of trust in automated agents,'' \emph{Frontiers in human
  neuroscience}, vol.~12, p. 309, 2018.

\bibitem{lu2020modeling}
Y.~Lu and N.~Sarter, ``Modeling and inferring human trust in automation based
  on real-time eye tracking data,'' in \emph{Proceedings of the Human Factors
  and Ergonomics Society Annual Meeting}, vol.~64, no.~1.\hskip 1em plus 0.5em
  minus 0.4em\relax SAGE Publications Sage CA: Los Angeles, CA, 2020, pp.
  344--348.

\bibitem{thropp2018calibrating}
J.~E. Thropp, T.~Oron-Gilad, J.~L. Szalma, and P.~A. Hancock, ``Calibrating
  adaptable automation to individuals,'' \emph{IEEE Transactions on
  Human-Machine Systems}, vol.~48, no.~6, pp. 691--701, 2018.

\bibitem{azevedo2020real}
H.~Azevedo-Sa, S.~K. Jayaraman, C.~T. Esterwood, X.~J. Yang, L.~P. Robert, and
  D.~M. Tilbury, ``Real-time estimation of drivers’ trust in automated
  driving systems,'' \emph{International Journal of Social Robotics}, pp.
  1--17, 2020.

\bibitem{Schaefer:2016hp}
\BIBentryALTinterwordspacing
K.~E. Schaefer, J.~Y.~C. Chen, J.~L. Szalma, and P.~A. Hancock, ``A
  {Meta}-{Analysis} of {Factors} {Influencing} the {Development} of {Trust} in
  {Automation},'' \emph{Human Factors: The Journal of the Human Factors and
  Ergonomics Society}, vol.~58, no.~3, pp. 377--400, Mar. 2016, iSBN:
  0018720816634. [Online]. Available:
  \url{http://journals.sagepub.com/doi/10.1177/0018720816634228}
\BIBentrySTDinterwordspacing

\bibitem{hoff2015}
K.~A. Hoff and M.~Bashir, ``Trust in automation: Integrating empirical evidence
  on factors that influence trust,'' \emph{Human factors}, vol.~57, no.~3, pp.
  407--434, 2015.

\bibitem{ayoub2019}
J.~Ayoub, F.~Zhou, S.~Bao, and X.~J. Yang, ``From manual driving to automated
  driving: A review of 10 years of autoui,'' in \emph{Proceedings of the 11th
  international conference on automotive user interfaces and interactive
  vehicular applications}, 2019, pp. 70--90.

\bibitem{zhang2019}
T.~Zhang, D.~Tao, X.~Qu, X.~Zhang, R.~Lin, and W.~Zhang, ``The roles of initial
  trust and perceived risk in public’s acceptance of automated vehicles,''
  \emph{Transportation research part C: emerging technologies}, vol.~98, pp.
  207--220, 2019.

\bibitem{zmud2016}
J.~Zmud, I.~N. Sener, J.~Wagner \emph{et~al.}, ``Consumer acceptance and travel
  behavior: impacts of automated vehicles,'' Texas A\&M Transportation
  Institute, Tech. Rep., 2016.

\bibitem{dzindolet2003}
M.~T. Dzindolet, S.~A. Peterson, R.~A. Pomranky, L.~G. Pierce, and H.~P. Beck,
  ``The role of trust in automation reliance,'' \emph{International journal of
  human-computer studies}, vol.~58, no.~6, pp. 697--718, 2003.

\bibitem{luo2020trust}
R.~Luo, J.~Chu, and X.~J. Yang, ``Trust dynamics in human-av (automated
  vehicle) interaction,'' in \emph{Extended Abstracts of the 2020 CHI
  Conference on Human Factors in Computing Systems}, 2020, pp. 1--7.

\bibitem{okamura2020}
K.~Okamura and S.~Yamada, ``Adaptive trust calibration for human-ai
  collaboration,'' \emph{PloS one}, vol.~15, no.~2, p. e0229132, 2020.

\bibitem{hergeth2016}
S.~Hergeth, L.~Lorenz, R.~Vilimek, and J.~F. Krems, ``Keep your scanners
  peeled: Gaze behavior as a measure of automation trust during highly
  automated driving,'' \emph{Human factors}, vol.~58, no.~3, pp. 509--519,
  2016.

\bibitem{walker2019}
F.~Walker, J.~Wang, M.~Martens, and W.~Verwey, ``Gaze behaviour and
  electrodermal activity: Objective measures of drivers’ trust in automated
  vehicles,'' \emph{Transportation research part F: traffic psychology and
  behaviour}, vol.~64, pp. 401--412, 2019.

\bibitem{akash2018}
K.~Akash, W.-L. Hu, N.~Jain, and T.~Reid, ``A classification model for sensing
  human trust in machines using eeg and gsr,'' \emph{ACM Transactions on
  Interactive Intelligent Systems (TiiS)}, vol.~8, no.~4, pp. 1--20, 2018.

\bibitem{vachon2014eye}
F.~Vachon and S.~Tremblay, ``What eye tracking can reveal about dynamic
  decision-making,'' \emph{Advances in cognitive engineering and
  neuroergonomics}, vol.~11, pp. 157--165, 2014.

\bibitem{zhou2021using}
F.~Zhou, X.~J. Yang, and J.~C. de~Winter, ``Using eye-tracking data to predict
  situation awareness in real time during takeover transitions in conditionally
  automated driving,'' \emph{IEEE Transactions on Intelligent Transportation
  Systems}, 2021.

\bibitem{hergeth2016keep}
S.~Hergeth, L.~Lorenz, R.~Vilimek, and J.~F. Krems, ``Keep your scanners
  peeled: Gaze behavior as a measure of automation trust during highly
  automated driving,'' \emph{Human factors}, vol.~58, no.~3, pp. 509--519,
  2016.

\bibitem{lu2019eye}
Y.~Lu and N.~Sarter, ``Eye tracking: a process-oriented method for inferring
  trust in automation as a function of priming and system reliability,''
  \emph{IEEE Transactions on Human-Machine Systems}, vol.~49, no.~6, pp.
  560--568, 2019.

\bibitem{zhou2011affect}
F.~Zhou, X.~Qu, M.~G. Helander, and J.~R. Jiao, ``Affect prediction from
  physiological measures via visual stimuli,'' \emph{International Journal of
  Human-Computer Studies}, vol.~69, no.~12, pp. 801--819, 2011.

\bibitem{hu2016real}
W.-L. Hu, K.~Akash, N.~Jain, and T.~Reid, ``Real-time sensing of trust in
  human-machine interactions,'' \emph{IFAC-PapersOnLine}, vol.~49, no.~32, pp.
  48--53, 2016.

\bibitem{khawaji2015using}
A.~Khawaji, J.~Zhou, F.~Chen, and N.~Marcus, ``Using galvanic skin response
  (gsr) to measure trust and cognitive load in the text-chat environment,'' in
  \emph{Proceedings of the 33rd Annual ACM Conference Extended Abstracts on
  Human Factors in Computing Systems}, 2015, pp. 1989--1994.

\bibitem{held2021heart}
J.~Held, A.~V{\^\i}sl{\u{a}}, C.~Wolfer, N.~Messerli-B{\"u}rgy, and
  C.~Fl{\"u}ckiger, ``Heart rate variability change during a stressful
  cognitive task in individuals with anxiety and control participants,''
  \emph{BMC psychology}, vol.~9, no.~1, pp. 1--8, 2021.

\bibitem{zhou2022predicting}
F.~Zhou, A.~Alsaid, M.~Blommer, R.~Curry, R.~Swaminathan, D.~Kochhar,
  W.~Talamonti, and L.~Tijerina, ``Predicting driver fatigue in monotonous
  automated driving with explanation using gpboost and shap,''
  \emph{International Journal of Human--Computer Interaction}, vol.~38, no.~8,
  pp. 719--729, 2022.

\bibitem{zhou2020driver}
F.~Zhou, A.~Alsaid, M.~Blommer, R.~Curry, R.~Swaminathan, D.~Kochhar,
  W.~Talamonti, L.~Tijerina, and B.~Lei, ``Driver fatigue transition prediction
  in highly automated driving using physiological features,'' \emph{Expert
  Systems with Applications}, p. 113204, 2020.

\bibitem{thayer2005psychosomatics}
J.~F. Thayer and J.~F. Brosschot, ``Psychosomatics and psychopathology: looking
  up and down from the brain,'' \emph{Psychoneuroendocrinology}, vol.~30,
  no.~10, pp. 1050--1058, 2005.

\bibitem{desai2013}
M.~Desai, P.~Kaniarasu, M.~Medvedev, A.~Steinfeld, and H.~Yanco, ``Impact of
  robot failures and feedback on real-time trust,'' in \emph{2013 8th ACM/IEEE
  International Conference on Human-Robot Interaction (HRI)}.\hskip 1em plus
  0.5em minus 0.4em\relax IEEE, 2013, pp. 251--258.

\bibitem{pop2015individual}
V.~L. Pop, A.~Shrewsbury, and F.~T. Durso, ``Individual differences in the
  calibration of trust in automation,'' \emph{Human factors}, vol.~57, no.~4,
  pp. 545--556, 2015.

\bibitem{benedek2010continuous}
M.~Benedek and C.~Kaernbach, ``A continuous measure of phasic electrodermal
  activity,'' \emph{Journal of neuroscience methods}, vol. 190, no.~1, pp.
  80--91, 2010.

\bibitem{du2020psychophysiological}
N.~Du, X.~J. Yang, and F.~Zhou, ``Psychophysiological responses to takeover
  requests in conditionally automated driving,'' \emph{Accident Analysis \&
  Prevention}, vol. 148, p. 105804, 2020.

\bibitem{du2020predicting}
N.~Du, F.~Zhou, E.~M. Pulver, D.~M. Tilbury, L.~P. Robert, A.~K. Pradhan, and
  X.~J. Yang, ``Predicting driver takeover performance in conditionally
  automated driving,'' \emph{Accident Analysis \& Prevention}, vol. 148, p.
  105748, 2020.

\bibitem{chen2016xgboost}
T.~Chen and C.~Guestrin, ``Xgboost: A scalable tree boosting system,'' in
  \emph{Proceedings of the 22nd acm sigkdd international conference on
  knowledge discovery and data mining}, 2016, pp. 785--794.

\bibitem{ayoub2022predicting}
J.~Ayoub, N.~Du, X.~J. Yang, and F.~Zhou, ``Predicting driver takeover time in
  conditionally automated driving,'' \emph{IEEE Transactions on Intelligent
  Transportation Systems}, 2022.

\bibitem{ayoub2022cause}
J.~Ayoub, Z.~Wang, M.~Li, H.~Guo, R.~Sherony, S.~Bao, and F.~Zhou,
  ``Cause-and-effect analysis of adas: A comparison study between literature
  review and complaint data,'' in \emph{Proceedings of the 14th International
  Conference on Automotive User Interfaces and Interactive Vehicular
  Applications}, 2022, pp. 139--149.

\bibitem{libby1973pupillary}
W.~L. Libby~Jr, B.~C. Lacey, and J.~I. Lacey, ``Pupillary and cardiac activity
  during visual attention,'' \emph{Psychophysiology}, vol.~10, no.~3, pp.
  270--294, 1973.

\bibitem{liu2022heart}
K.~Y. Liu, T.~Elliott, M.~Knowles, and R.~Howard, ``Heart rate variability in
  relation to cognition and behavior in neurodegenerative diseases: A
  systematic review and meta-analysis,'' \emph{Ageing research reviews},
  vol.~73, p. 101539, 2022.

\end{thebibliography}
\begin{IEEEbiography}[{\includegraphics[width=1in,height=1.55in,clip,keepaspectratio]{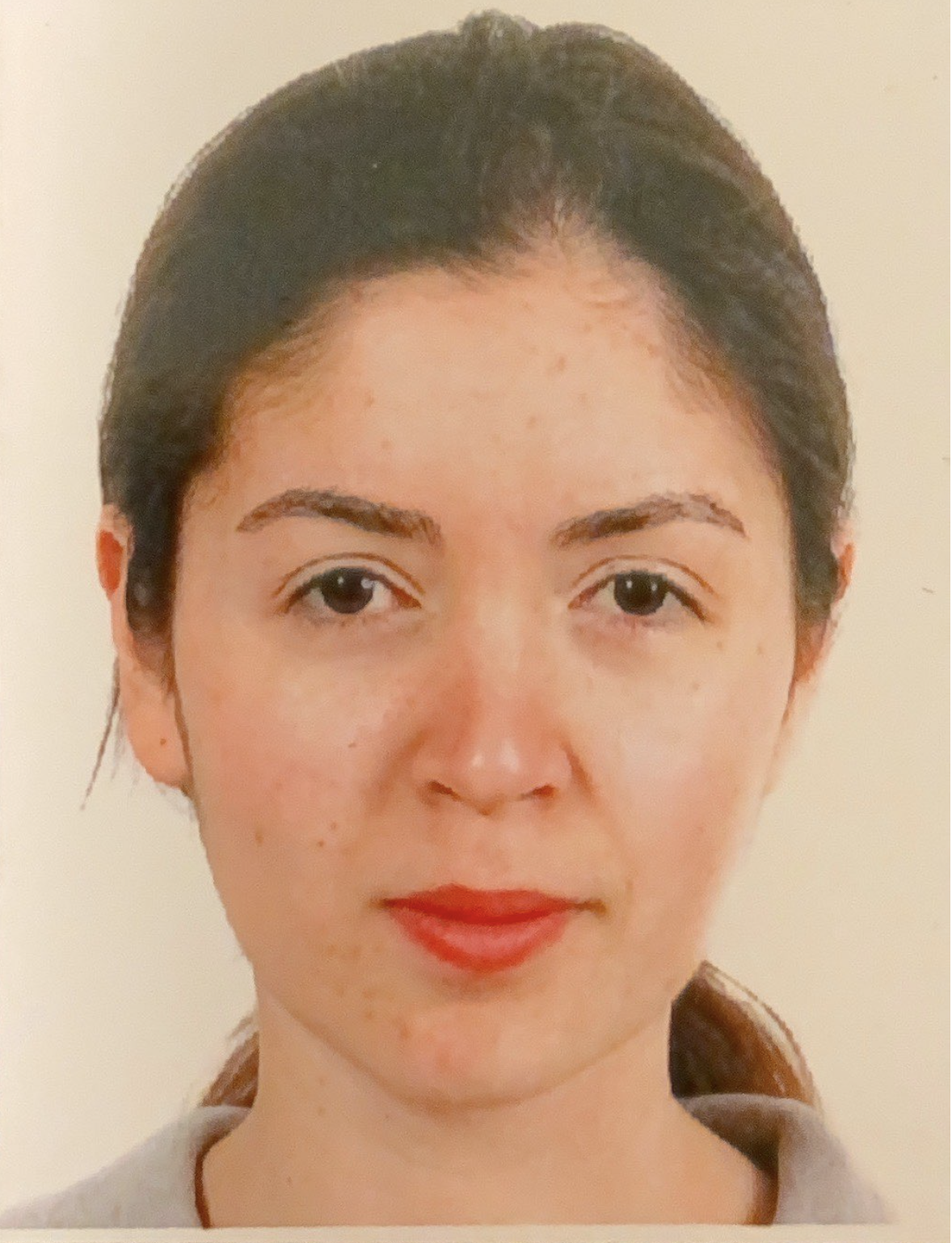}}]{Jackie Ayoub received her B.E. degree in mechanical engineering from Notre Dame University, Lebanon, in 2016 and her master degree in Industrial and Systems Engineering from University of Michigan, Dearborn, in 2017. She received her Ph.D. in Industrial and Systems Engineering in the University of Michigan, Dearborn, in 2022. She is currently a data scientist and human factor engineer at Honda Research Institute, Detroit. Her main research interests include human-computer interaction, human factors and ergonomics, and sentiment analysis.}
\vspace{-20pt}
\end{IEEEbiography}

\begin{IEEEbiography}[{\includegraphics[width=1in,height=1.55in,clip,keepaspectratio]{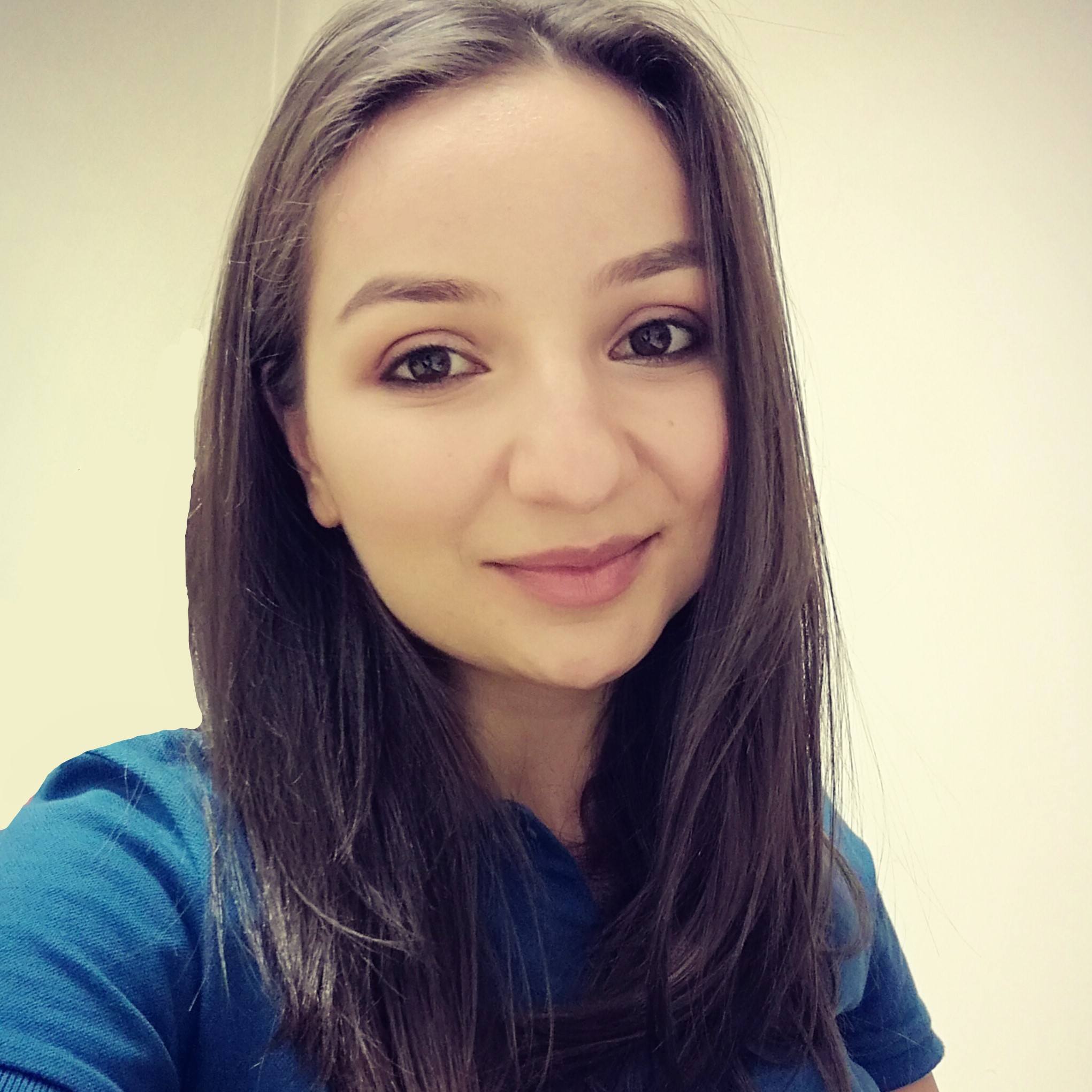}}]{Lilit Avetisyan received her B.E. degree in 2017 and MS degree in 2019 in Information Security from the National Polytechnic University of Armenia. She is currently pursuing her Ph.D. degree in Industrial and Systems Engineering at the University of Michigan, Dearborn. Her main research interests include human-computer interaction, explainable artificial intelligence and situation awareness.}
\vspace{-20pt}
\end{IEEEbiography}

\begin{IEEEbiography}[{\includegraphics[width=1in,height=1.55in,clip,keepaspectratio]{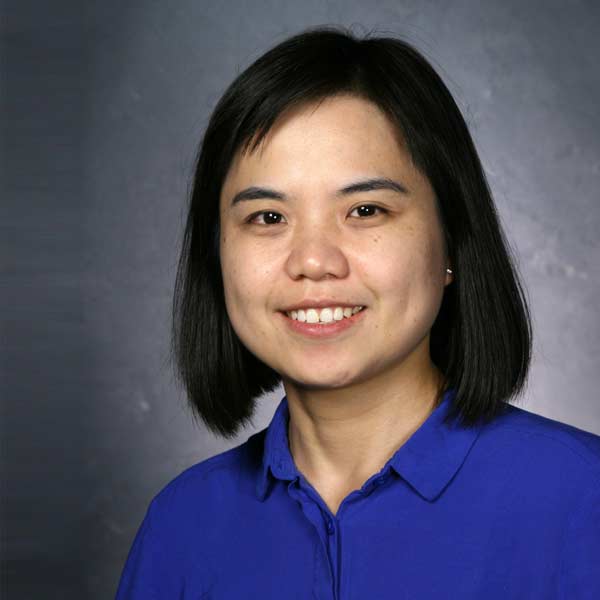}}]{X. Jessie Yang is an Assistant Professor in the Department of Industrial and Operations Engineering, University of Michigan, Ann Arbor. She earned a PhD in Mechanical and Aerospace Engineering (Human Factors) from Nanyang Technological University, Singapore. Dr. Yang’s research include human-autonomy interaction, human factors in high-risk industries and user experience design.}
\vspace{-20pt}
\end{IEEEbiography}

\begin{IEEEbiography}[{\includegraphics[width=1in,height=1.55in,clip,keepaspectratio]{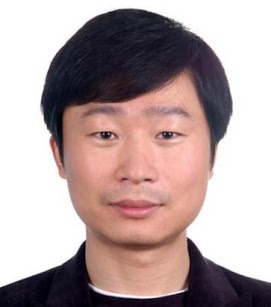}}]{Dr Feng Zhou received the Ph.D. degree in Human Factors Engineering from Nanyang Technological University, Singapore, in 2011 and Ph.D. degree in Mechanical Engineering from Gatech Tech in 2014. He was a Research Scientist at MediaScience, Austin TX, from 2015 to 2017. He is currently an Assistant Professor with the Department of Industrial and Manufacturing Systems Engineering, University of Michigan, Dearborn. His main research interests include human factors, human-computer interaction, engineering design, and human-centered design.}
\end{IEEEbiography}



\end{document}